\documentstyle[onecolumn,epsf,mnras_cite]{mn}
\input{epsf}
\newcommand{\ba}{\begin{eqnarray}}
\newcommand{\ea}{\end{eqnarray}}
\newcommand{\be}{\begin{equation}}
\newcommand{\ee}{\end{equation}}
\newcommand{\nn}{\nonumber \\}
\newcommand{\vk}{{\bf{k}}}
\newcommand{\vkone}{{\bf{k}_1}}
\newcommand{\vktwo}{{\bf{k}_2}}
\newcommand{\vkthree}{{\bf{k}_3}}

\newcommand{\vr}{{\bf{r}}}
\newcommand{\vx}{{\bf{x}}}
\newcommand{\ls}{\mathrel{\raise1.16pt\hbox{$<$}\kern-7.0pt 
\lower3.06pt\hbox{{$\scriptstyle \sim$}}}}         
\newcommand{\gs}{\mathrel{\raise1.16pt\hbox{$>$}\kern-7.0pt 
\lower3.06pt\hbox{{$\scriptstyle \sim$}}}}         

\long\def\comment#1{}

\def\fun#1#2{\lower3.6pt\vbox{\baselineskip0pt\lineskip.9pt
  \ialign{$\mathsurround=0pt#1\hfil##\hfil$\crcr#2\crcr\sim\crcr}}}

\newcommand{\vkappa}{\mbox{\boldmath $\kappa$}}
\newcommand{\vtheta}{\mbox{\boldmath $\theta$}}
\title{Projected bispectrum in spherical harmonics and its application to 
angular galaxy catalogues}
\author[Licia Verde, Alan F. Heavens, Sabino Matarrese]
{Licia Verde$^{1}$, Alan F. Heavens$^{1}$, Sabino Matarrese$^{2,3}$\\
$^{1}$ Institute for Astronomy, University of Edinburgh, Royal Observatory,
Blackford Hill, Edinburgh EH9 3HJ, United Kingdom\\
$^{2}$ Dipartimento di Fisica {\em Galileo Galilei}, Universit\`{a} di
Padova, via Marzolo 8, I-35131 Padova, Italy\\
$^{3}$ Max-Planck-Institut f\"ur Astrophysik, 
Karl-Schwarzschild-Strasse 1, D-85748 Garching, Germany. }
\begin{document}
\maketitle
\begin{abstract}
We present a theoretical and exact analysis of the bispectrum of projected
galaxy catalogues. The result can be generalized to evaluate the projection 
in spherical harmonics of any 3D bispectrum and therefore has
applications to cosmic microwave background and gravitational lensing studies.

By expanding the 2D distribution of galaxies on
the sky in spherical harmonics, we show how the 3-point function of
the coefficients can be used in principle to determine the bias
parameter of the galaxy sample.  If this can be achieved, it would
allow a lifting of the degeneracy between the bias and the
matter density parameter of the Universe which occurs in linear
analysis of 3D galaxy catalogues.  In previous papers we have shown
how a similar analysis can be done in three dimensions, and we show
here through an error analysis and by implementing the method on a
simulated projected catalogue that ongoing three-dimensional galaxy
redshift surveys (even with all the additional uncertainties introduced by partial sky coverage, redshift-space distortions and smaller numbers) will do far better than all-sky
projected catalogues with similar selection function.
\end{abstract}


\section{Introduction}

The clustering of mass in the Universe is an important fossil record
of the early perturbations which gave rise to large-scale structure
today.  Knowledge of the mass clustering puts powerful constraints on
the quantity and properties of dark matter in the Universe, and the
generation mechanism for the perturbations.  Most of our knowledge of
the mass clustering is, however, indirect, coming principally from the
distribution of galaxies.  A major obstacle in interpretation is
therefore the uncertain relationship between galaxy and mass
clustering - a relationship which is conventionally quantified by the
`bias parameter', $b$.  In particular, attempts based on linear
perturbation theory to measure the density parameter of the Universe,
$\Omega_0$, through peculiar velocity or redshift-distortion studies,
yield only the degenerate combination $\beta=\Omega_0^{0.6}/b$, making
it impossible to determine $\Omega_0$ without determining $b$.  The
degeneracy can be lifted by going to second order in perturbation
theory, and this can be achieved most elegantly by studying the
bispectrum, which is the three-point function in Fourier (or spherical
harmonic) space.  A major positive feature of the bispectrum method is
that it can provide error bars on the desired parameters.  The method
works because gravitational instability leads to a density field which
is progressively more skewed to high densities as it develops, and
this skewness appears as a non-zero bispectrum.  This behaviour can
also be mimicked by biasing, if the galaxy density field is a local,
nonlinear function of the underlying mass density field.  This
possibility must be dealt with.  The two effects can, however, be
separated by the use of information about the shape of the structures:
in essence the effect of biasing is to shift iso-density contours up
(or down), while maintaining the shape of the contour; gravitational
evolution, instead, changes the shape, usually leading to flattening
of collapsing structures (e.g. Zel'dovich pancakes).

\scite{Fry94} recognized the role the bispectrum could play in
determining the bias parameter, and \scite{MVH97} (hereafter MVH97)
and \scite{VHMM98} have turned the idea into a practical proposition
for 3D galaxy redshift surveys by including analysis of selection
functions, shot noise, and redshift distortions (see also
\pcite{SCF98}).  The latter is potentially a serious problem for 3D
surveys, as the signal for bias comes from mildly nonlinear scales,
where the redshift distortions are not trivial to analyze.  However,
experiments on simulated catalogues \cite{VHMM98} show that the method
is successful.  Note however that the theory has been developed only
in the `distant-observer approximation' (see e.g. \pcite{Kaiser92}), and is
applicable to relatively deep surveys such as the Anglo-Australian
two-degree field galaxy redshift survey (\pcite{colles96,Colles99})
and the Sloan Digital Sky Survey (\pcite{SDSS}).  Shallow surveys such
as the IRAS PSCz should be analysed in spherical coordinates
(cf Taylor \& Heavens 1995; Tadros et al. 1999 for the 
power spectrum), and suffer from high shot noise, so cannot be usefully used
for bispectrum analysis based on a Fourier expansion (see also 
\pcite{Bharadwaj}).

The absence of suitable existing 3D surveys prompts us to consider
whether the bias might be extracted from a projected galaxy catalogue,
such as the APM galaxy survey (\pcite{maddoxetal90,Lov92,JonAPMcat}).
With only angular positions, the information is more limited, but the
survey is not complicated by redshift distortions, and contains a
large number $\sim 10^6$ of galaxies.  The DPOSS catalogue
(\pcite{DPOSS}) will be even larger, with 50 million galaxies and
nearly all-sky. 
There are two important caveats to keep in mind: first, to have a
measurement of $\Omega_0$ we need to be able to measure the $\beta$ parameter
and the linear bias parameter. It is not possible to extract the $\beta$
parameter from a two-dimensional survey, $\beta$ will need to be determined
from a -{\em different}- three-dimensional survey. The selection criteria will
necessarily be different for different catalogues, and so will be the galaxy
population selected. Since different galaxy populations can have different bias
with respect to the underlying dark matter distribution, some care needs to be
taken in the interpretation of the final result i.e. the value for $\Omega_0$.
The other caveat concerns the effect of the evolution along the line of sight
(also referred to as the {\it light-cone effect}). This is due to the fact
that galaxy clustering evolves gravitationally with time along the line of
sight and depends on the (unknown) cosmology. For shallow surveys such as the
APM, this effect is smaller or comparable to the cosmic variance, in what follows we will neglect this effect for this
reason. However, for deeper
surveys, this effect needs to be properly taken into account.  Assuming these
issues can be dealt with, the key requirement is to  obtain an expression
for the projected bispectrum given an analytical formula for the spatial one.
An expression for the projected bispectrum in the small angle approximation
has been presented by \scite{BKJ99}.
However this might not be a good approximation for the bispectrum if the
sample is close enough to the observer or if the scales under analysis are large. In fact it is not known {\it a priori} whether, for the bispectrum,  the small angle approximation is valid on large enough
scales for the second order perturbation theory to hold:
it is necessary to obtain an exact expression for the projected
bispectrum using spherical harmonics expansion. Only then it will be possible to
test the limit of validity for the small-angle approximation bispectrum (Verde
et al. 2000).

In this paper, we develop the theory for projected catalogues in a
full treatment.  The resulting expression for the spherical harmonic projected
bispectrum can straightforwardly be
applied to gravitational lensing and to cosmic microwave background (CMB)
studies, for comparison with observations such as the claimed detection by \scite{FMG98}.

 In Section 2 we expand the sky density of galaxies in
spherical harmonics with coefficients $a_\ell^m$, and compute an
explicit expression for the bispectrum $\langle a_{\ell_1}^{m_1}
a_{\ell_2}^{m_2} a_{\ell_3}^{m_3}
\rangle$ accurate to second-order in perturbation theory.  In
particular, we show how this quantity depends on the bias parameter.
We present in Section 3 an error analysis specific to the second-order
perturbation theory bispectrum, which shows the expected
uncertainty in the derived bias parameter, and test on a numerical
simulation.  In the Appendices, we detail asymptotic results which are 
useful for high-$\ell$ spherical harmonics.  The main conclusion of
 large-scale structure application in this paper is that 3D large scale
structure surveys (even with small sky coverage, smaller numbers, and the complications of
redshift-space distortion, shot noise etc.) will do far better than all-sky 
projected catalogues for the purpose of measuring the bias parameter. 
 However the mathematics developed for this purpose has much wider applications: with
appropriate radial weight functions, the analysis can be applied to the CMB
bispectrum induced by lensing, Sunyaev-Zel'dovich effect, the integrated
Sachs-Wolfe effect  or foreground point sources, and to gravitational lenses studies.

\section{Projected Bispectrum in Spherical Harmonics}

Let the projected galaxy density field be $n(\Omega)$, where $\Omega$
represents angular positions in the sky.  If the three-dimensional
galaxy density field is $\rho(\vr)$ (with mean $\overline{\rho}$) and
the selection function is $\psi(r)$, the projected density is

\be
n(\Omega)d\Omega =\left(\int dr\, r^2 \rho(\vr)\psi(r)\right)d\Omega.
\ee

We expand the projected density in spherical harmonics (see Appendix A
for definitions)
\ba
a^m_{\ell} & \equiv &\frac{1}{\overline{n}}\int d\Omega\,
n(\Omega)Y_{\ell}^m(\Omega)\nn
& =&\frac{\overline{\rho}}{\overline{n}}\int
d\Omega dr \, r^2 \delta(r)\psi(r) Y_{\ell}^m(\Omega)\mbox{  for  ${\ell}\neq 0$ }
\label{blm}
\ea
where $\delta \equiv (\rho-\overline{\rho})/\overline{\rho}$ is the
fractional overdensity in galaxies.  The average surface density is
\be
\overline{n}=\int dr\,r^2\overline{\rho}\psi(r).
\ee
and is inserted  in the transform for convenience.  The three-point function of the
coefficients may be factorised by isotropy (e.g. \pcite{Luo94}) into
\be
\langle a_{\ell_1}^{m_1} a_{\ell_2}^{m_2}  a_{\ell_3}^{m_3}\rangle=B_{\ell_1\ell_2\ell_3}\left(^{\ell_1\;\;\;\ell_2\;\;\;\ell_3}_{m_1m_2m_3} \right)
\label{eq:2dbispluo}
\ee
where $\left(^{\ell_1\;\;\;\ell_2\;\;\;\ell_3}_{m_1m_2m_3} \right)$ is the
Wigner 3J symbol. We refer to $B_{\ell_1\ell_2\ell_3}$ as the
angular bispectrum.  From general considerations about rotational
invariance of the quantity $\langle
a_{\ell_1}^{m_1}a_{\ell_2}^{m_2}a_{\ell_3}^{m_3}
\rangle $ the indices $\ell_i, m_i$ for $i=1,2,3$ must satisfy the
following conditions:
\begin{itemize}
\item[{\it (i)}] $\ell_j +\ell_k\geq\ell_1 \geq \mid \ell_j- \ell_k \mid$ (triangle rule)
\item[{\it (ii)}] $\ell_1 +\ell_2+\ell_3=$ even
\item[{\it (iii)}] $m_1+m_2+m_3=0$.
\end{itemize}
The presence of the 3J symbol ensures that these conditions are
satisfied. 

In order to be able
to extract the bias parameter from projected catalogues, the effect of the
projection in the configuration dependence of the bispectrum needs to be
understood (\pcite{FryThomas99}).

To do so, we compute the angular bispectrum in terms of the 3D
bispectrum, $B(\vk_1,\vk_2,\vk_3)$ defined by  $\langle \delta_{\vk1}
\delta_{\vk2}\delta_{\vk3}\rangle=(2\pi)^3 B(\vk_1,\vk_2,\vk_3) 
\delta^D(\vk_1+\vk_2+\vk_3)$, where the Fourier transform of
$\delta$ is $\delta_\vk \equiv \int d^3\vr ~\delta(\vr) \exp(i\vk\cdot
\vr)$, and $\delta^{D}$ denotes the Dirac delta function.

We proceed from (\ref{blm}):
\be
\langle a_{\ell_1}^{m_1} a_{\ell_2}^{m_2}  a_{\ell_3}^{m_3}\rangle=
\left(\frac{\overline{\rho}}{\overline{n}}\right)^3\int
d\Omega_1 d\Omega_2 d\Omega_3 dr_1 dr_2 dr_3 r_1^2 r_2^2r_3^2
\psi_1 \psi_2 \psi_3
 \langle\delta(\vr_1)\delta(\vr_2)\delta(\vr_3) \rangle
 Y_{\ell_1}^{m_1}(\Omega_1)Y_{\ell_2}^{m_2}(\Omega_2)Y_{\ell_3}^{m_3}
(\Omega_3) \;.
\ee
The 3D three-point function (in real space) is related to the 3D
bispectrum by
\be
\langle \delta(\vr_1)\delta(\vr_2)\delta(\vr_3) \rangle=
\frac{1}{(2\pi)^6}\int d^3\vk_1 d^3\vk_2 d^3\vk_3 B(\vk_1,\vk_2,\vk_3)e^{i(\vk_1\cdot
\vr_1+\vk_2\cdot \vr_2+ \vk_3\cdot
\vr_3)}\delta^{D}(\vk_1+\vk_2+\vk_3) \;.
\ee
We then define the quantity:
\be
I(\vr_1,\vr_2,\vr_3) 
\equiv\int_0^{\infty}dk_1dk_2dk_3k_1^2k_2^2k_3^2\int_{4\pi}d\Omega_{k_1}
d\Omega_{k_2} d\Omega_{k_3} B(\vkone,\vktwo,\vkthree)
e^{i(\vkone\cdot\vr_1+\vktwo\cdot\vr_2+\vkthree\cdot\vr_3)}
\delta^D(\vkone+\vktwo+\vkthree)
\label{eq.I123}
\ee
because we will later expand the exponential in spherical harmonics and
perform the angular integrations in (\ref{eq.I123}) explicitly.

In second order perturbation theory the bispectrum is:
\be
B(\vkone,\vktwo,\vkthree)= {\cal K}(\vkone,\vktwo)P(k_1)P(k_2)+cyc. \;,
\label{eq:2OPTbisp}
\ee
where the shape-dependent factors ${\cal K}$ can be found in
\scite{Fry84}, \scite{CLMM95} and MVH97.  The dependence of ${\cal
K}$ on the cosmology in negligible (e.g. \scite{KB99}
and references therein), so in what follows we
assume an Einstein-de Sitter Universe.  The factors are, however,
dependent on the biasing model assumed.  If we take a local biasing
model, then for consistency with second-order perturbation theory, we
expand in a Taylor series the galaxy overdensity to second-order in
the matter overdensity $\delta_m$:
\be
\delta(\vx)=b_1\delta_m(\vx)+{1\over 2}b_2\delta_m(\vx)^2,
\label{eq:nlbias}
\ee
(a constant term $b_0$ is irrelevant except at $\vk={\bf 0}$ and is
ignored). Here $b_1$ is the linear bias parameter and $b_2$ is the
quadratic bias parameter. The linear bias parameter $b$ that appears
in the definition of $\beta$ and that is needed to recover $\Omega_0$,
is $b = b_1$ on large scales, under fairly general conditions
(\pcite{HMV98}).  Note that we take the bias function to be
deterministic, not stochastic (cf
\pcite{CO92,DL99,TegBrom99,Matsu99,Somervilleetal99}); 
it has been shown (\pcite{Taruyaetal99})
that the effect of stochastic bias on the bispectrum is very similar
to that of nonlinear bias (Eq. \ref{eq:nlbias}). 
This formalism might be straightforwardly extended to the case when the
bias process operates in Lagrangian, rather than Eulerian, space 
(\pcite{Catelanetal98}).  

With these assumptions, a typical cyclical term can be written
\be
{\cal
K}(\vkone,\vktwo)=A_0+A_1\cos(\theta_{12})+A_2\cos^2(\theta_{12})
\ee
where $\theta_{12}$ denotes the angle between $\vkone$ and $\vktwo$, and
\ba
A_0  &=&  \frac{10}{7}c_1+c_2 \nn
A_1  &=&  c_1\left(\frac{k_1}{k_2}+\frac{k_2}{k_1}\right) \nn
A_2  &=& \frac{4}{7}c_1,
\label{eq.A0A1A2}
\ea
and $c_1=1/b_1$; $c_2=b_2/b_1^2$.  Through these relations, the
projected bispectrum will depend on the bias parameters $b_1$ and $b_2$.

Using this, we will now calculate the theoretical expression for the
projected bispectrum in spherical harmonics in the mildly nonlinear
regime.  With substitution (\ref{eq:2OPTbisp}) we find
\be
I(\vr_1,\vr_2,\vr_3)\equiv I_{12}+I_{23}+I_{13}
\ee

Using the properties of spherical harmonics in Appendix A [equation
(\ref{eq.expspharm}), the orthogonality relation, (\ref{eq.spharmorth})
and (\ref{eq.spharmconj})], we obtain that a typical cyclical term is:

\ba
I_{12}&=&(4\pi)^4\int_0^{\infty}d^3\vk_1d^3\vk_2
{\cal K}(\vk_1,\vk_2)P(k_1)P(k_2)\times \nn
& &\sum_{\ell_1^{\prime}m_1^{\prime}}i^{\ell_1^{\prime}}j_{\ell_1^{\prime}}(k_1r_1)Y_{\ell_1^{\prime}}^{*m_1^{\prime}}(\Omega_{k_1})Y^{m_1^{\prime}}_{\ell_1^{\prime}}(\Omega_{r_1})
\sum_{\ell_2^{\prime}m_2^{\prime}}i^{\ell_2^{\prime}}j_{\ell_2^{\prime}}(k_2r_2)Y_{\ell_2^{\prime}}^{*m_2^{\prime}}(\Omega_{k_2})Y^{m_2^{\prime}}_{\ell_2^{\prime}}(\Omega_{r_2})\times \nn
& & \sum_{L_1 n_1}i^{L_1}j_{L_1}(k_1r_3)Y_{L_1}^{*n_1}(-\Omega_{k_1}) Y^{n_1}_{L_1}(\Omega_{r_3}) \sum_{L_2n_2}i^{L_2}j_{L_2}(k_2r_3)Y_{L_2}^{*n_2}(-\Omega_{k_2}) Y^{n_2}_{L_2}(\Omega_{r_3}).
\label{eq:I12} 
\ea 

We can now write 

\be 
\int_{4\pi}d\Omega Y^{*m_1}_{\ell_1}(\Omega )Y^{m_2}_{\ell_2}(\Omega
)Y^{m_3}_{\ell_3}(\Omega)= {\cal H}^{m_1m_2m_3}_{\ell_1\; \ell_2\; \ell_3} \;,
\ee
which can be expressed
in term of Clebsch-Gordan coefficients, and is non-zero only if
the following symmetry conditions are satisfied:
\begin{itemize}
\item $m_2+m_3=m_1$
\item $\ell_1+\ell_2+\ell_3=$ even
\item $\ell_1,\ell_2,\ell_3$ satisfy the triangle rule
\end{itemize}
From equation (\ref{eq:I12}) we thus obtain:
\be
\int_{4\pi}d\Omega_{r_1}d\Omega_{r_2}d\Omega_{r_3}
Y_{\ell_1}^{m_1}
(\Omega_{r_1})Y_{\ell_2}^{m_2}(\Omega_{r_2})Y_{\ell_3}^{m_3}(\Omega_{r_3})I_{12}=(4\pi)^4\int dk_1
dk_2k_1^2k_2^2P(k_1)P(k_2)F(r_1,r_2,r_3,k_1,k_2) \;,
\ee
where
\ba
 F(r_1,r_2,r_3,k_1,k_2)&=& \int d\Omega_{k_1}d\Omega_{k_2} \left(A_0
 +A_1\cos\theta_{12}+A_2\cos^2\theta_{12} \right)i^{\ell_1+\ell_2}
 (-1)^{(m_1+m_2+m_3)}j_{\ell_1}(k_1r_1)j_{\ell_2}(k_2r_2)  \times \nn & &
 \sum_{\ell_{6,7}m_{6,7}}i^{\ell_6+\ell_7}j_{\ell_6}(k_1r_3)j_{\ell_7}(k_2r_3)Y^{*-m_1}_{\ell_1}(\Omega_{k_1})Y^{*-m_2}_{\ell_2}
 (\Omega_{k_2})Y^{*m_6}_{\ell_6}(-\Omega_{k_1})Y^{*m_7}_{\ell_7}(-\Omega_{k_2}){\cal H}^{-m_3m_6m_7}_{\ell_3\;\ell_6\;\ell_7}
\ea
and F can be written as $F_0+F_1+F_2$ where $F_0$ involves the term
$A_0$ etc.

The $F_0$ term is easily calculated: 

\be
F_0=A_0i^{2(\ell_1+\ell_2)}j_{\ell_1}(k_1r_1)j_{\ell_2}(k_2r_2)j_{\ell_1}(k_1r_3)j_{\ell_2}(k_2r_3)
{\cal H}_{\ell_3 \;\ell_1 \; \ell_2}^{-m_3 m_1 m_2} 
\ee 
and therefore satisfies the symmetry rules.

For $F_1$ and $F_2$ we exploit the fact that:
\be
\cos\theta_{12}=P_1(\cos\theta_{12}),
\ee
\be
\cos^2\theta_{12}=\frac{1}{3}\left[2 P_2(\cos\theta_{12})+P_0(\cos\theta_{12}) \right]
\ee
and use the addition theorem for spherical harmonics:
\be
P_n(\cos\theta_{12})=\frac{4 \pi}{2n+1}\sum_{m=-n}^nY_n^m(\Omega_{k_1})Y_n^{*m}(\Omega_{k_2}).
\ee
The $F_1$ term then becomes:
\be
F_1=A_1\frac{4\pi}{3}
i^{\ell_1+\ell_2}(-1)^{m_1+m_3}j_{\ell_1}(k_1r_1)j_{\ell_2}(k_2r_2)\sum_{\ell_6m_6 \ell_7m_7 M}
i^{\ell_6+\ell_7}j_{\ell_6}(k_1r_3)j_{\ell_7}(k_2k_3) {\cal H}^{-m_3 m_6
m_7}_{\ell_3 \; \ell_6 \;\ell_7}{\cal H}^{-m_1 -m_6 M}_{\ell_1 \; \ell_6
\;1}{\cal H}^{M m_2 -m_7 }_{1\; \ell_2\; \ell_7}.
\label{eq.F1}
\ee
It is easy to see that $m_6+m_7=-m_3$.  To demonstrate that the
symmetry conditions are all satisfied, consider the
following part of eq.(\ref{eq.F1}):
\be
\sum_{m_6m_7M}{\cal
H}^{-m_3 m_6 m_7}_{\ell_3 \; \ell_6 \; \ell_7}{\cal H}^{-m_1 -m_6 M}_{\ell_1
\;\ell_6 \; 1}{\cal H}^{M m_2 -m_7}_{1\; \ell_2 \; \ell_7} \;;
\ee
let's introduce a new
quantity $h^{m_1m_2m_3}_{\ell_1 \; \ell_2 \; \ell_3}$ that is symmetric for any
permutation of the columns $\left(^{m_i}_{\ell_i}\right)$. It is
clear that:
\be
{\cal
H}^{m_1m_2m_3}_{\ell_1 \; \ell_2 \; \ell_3}=(-1)^{m_1}h^{-m_1m_2m_3}_{\ell_1
\; \ell_2 \; \ell_3}.
\ee
The quantity $h^{m_1m_2m_3}_{\ell_1 \; \ell_2 \; \ell_3}$ can be written in terms of the $3J$ symbols:
\be
h^{m_1m_2m_3}_{\ell_1 \; \ell_2 \; \ell_3}=\sqrt{(2 \ell_1+1)(2\ell_2+1)(2\ell_3+1)/(4 \pi)}
\left(^{\ell_1\; \ell_2\;\ell_3}_{0\;\;\;0\;\;\;0}\right)
\left(^{\ell_1\;\;\;\ell_2\;\;\;\ell_3}_{m_1m_2m_3} \right).
\ee
Equation (\ref{eq.F1}) therefore contains the following multiplicative term:
\be
\sum_{m_6m_7M}(-1)^{-m3-m_1+M}\left(^{\ell_3\;\;\;\;\ell_6\;\;\;\;\ell_7}_{m_3\;m_6\;m_7}\right)
\left(^{\ell_1\;\;\;\;\ell_6\;\;\;\;1}_{m_1\;-m_6 \;M} \right)
\left(^{1\;\;\;\;\;\ell_2\;\;\;\;\;\;\ell_7}_{-M\;m_2\;-m_7} \right)=(-1)^{\ell_6+\ell_7+\ell}
\left(^{\ell_1\;\;\;\;\ell_2\;\;\;\;\ell_3}_{m_1\;m_2\;m_3} \right)
\left\{^{\ell_1\;\;\;\ell_2\;\;\;\ell_3}_{\ell_7\;\;\; \ell_6\;\;\;\; 1}\right\}
\label{eq.3j3j3j} \;,
\ee
where $\left\{^{\ell_1\;\;\;\ell_2\;\;\;\ell_3}_{\ell_7\;\;\; \ell_6\;\;\;\;
1}\right\}$ denotes the 6J symbol.
In the last equality we used  eq. (4.88) of \scite{Sobelman}.
The properties of the 3J symbol ensure that the $F_1$ term
satisfies the symmetry conditions.

Similarly for the $F_2 $ term we obtain
\ba
F_2 & = & A_2 i^{\ell_1+\ell_2}(-1)^{m_1+m_3}\frac{4\pi}{3}
j_{\ell_1}(k_1r_1)j_{\ell_2}(k_2r_2)
\sum_{\ell_{6,7}m_{6,7}M}(-i)^{\ell_6+\ell_7}j_{\ell_6}(k_1r_3)j_{\ell_7}(k_2r_3) \times \nn
 & & {\cal H}^{-m3m_6 m_7}_{\ell_3\;\;\; \ell_6\;\;\; \ell_7}
\left[ \frac{2}{5}{\cal H}^{-m1 M -m_6}_{\ell_1\;\;\; 2\;\;\; \ell_6}{\cal H}^{M m_2
-m_7}_{2\;\;\; \ell_2\;\;\; \ell_7} + {\cal H}^{-m_1 0-m_6}_{\ell_1\;\;\;0\;\;\; \ell_6}
{\cal H}^{ 0 m_2 -m_7}_{0\;\;\; \ell_2\;\;\; \ell_7} \right] \;.
\ea
The second term in the square brackets  does not present any
problem, in fact it is nonzero only if $\ell_1=\ell_6$, $\ell_2=\ell_7$,
$m_1=m_6$, $m_2=m_7$, and satisfies the symmetry conditions.
Similar methods to those above complete the symmetry considerations.

Factorising the Wigner 3J symbol, and collecting terms together, we
find the expression for the angular bispectrum, as
a sum of cyclical permutations:
\be
B_{\ell_1 \ell_2\ell_3}={\cal B}_{12}+{\cal B}_{13}+{\cal B}_{23}
\label{eq:bisp1}
\ee
where, writing $\Psi_{\ell}(k)=\overline{\rho}\int dr r^2j_{\ell}(kr)\psi(r)$,
\ba
{\cal B}_{12}&=&\frac{1}{\overline{n}^3}\frac{16}{\pi}\sqrt{\frac{(2\ell_1+1)(2\ell_2+1)(2\ell_3+1)}{(4\pi)^3}}\int dk_1 dk_2 i^{\ell_1+\ell_2} k_1^2
k_2^2P(k_1)P(k_2)\Psi_{\ell_1}(k_1)\Psi_{\ell_2}(k_2) \times \nn
& &\!\!\!\!\!\!\!\!\!\sum_{\ell \ell_6\ell_7}i^{\ell_6+\ell_7}(-1)^{\ell} B_{\ell}(k_1,k_2)(2 \ell_6+1)(2\ell_7+1)\overline{\rho}\!\int dr r^2 \psi(r)j_{\ell_6}(k_1r)j_{\ell_7}(k_2r)
\left(^{\ell_1\;\;\ell_6\;\;\ell}_{0 \;\;\;0\;\;\; 0} \right)
\left(^{\ell_2\;\;\;\ell_7\;\;\ell}_{0\;\;\;\; 0\;\;\;\; 0} \right)
\left(^{\ell_3\;\;\ell_6\;\;\ell_7}_{0\;\;\;\; 0\;\;\;\; 0} \right)
\left\{^{\ell_1\;\ell_2\;\ell_3}_{\ell_7\;\ell_6\;\,\ell}
\right\}
\label{eq.finalprojbisp}
\ea
where $B_{\ell}(k_1,k_2)$ for $\ell=0,1,2$ are:

\ba
B_0(k_1,k_2)&=&  \frac{34}{21}c_1+c_2 \nn
B_1(k_1,k_2)&=&  c_1\left(\frac{k_1}{k_2}+\frac{k_2}{k_1}\right) \nn
B_2(k_1,k_2)&=& \frac{8}{21}c_1,
\label{eq.B0B1B2}
\ea
and the sum  $\sum_{\ell\ell_6\ell_7}$ extends over $\ell=0,1,2;$
$\ell_6=\ell_1-\ell.....\ell_1+\ell;$ 
$\ell_7=\ell_2-\ell.....\ell_2+\ell$.

The above expression can easily be generalized for any 3D bispectrum. In
fact, since  a) the bispectrum is non-zero only if the three $k$ vectors form a
triangle, b) the bispectrum does not depend on the spatial orientation of the
triangle (isotropy)  and c) a triangle is completely specified only by the
magnitude of two sides and the angle between them, the bispectrum can always
be expressed as a sum over three cyclical terms each involving only the
modulus of two $k$-vectors and the angle between them:

\be
B(\vk_1,\vk_2,\vk_3)={\cal F}(k_1,k_2,\theta_{12}).
\ee
Each of the cyclical terms  can therefore be expanded as:
\be
{\cal F}(k_1,k_2,\theta_{12})=P(k_1)P(k_2)\sum_{\ell=0}^nB_{\ell}(k_1,k_2)P_{\ell}(\cos\theta_{12})
\ee
where now $P(k_i)$ is an arbitrary function of $|\vk_i|$,  the
coefficients $B_{\ell}$ can depend on any combination of $|\vk_1|$ and
$|\vk_2|$ and the sum over $\ell$ should in principle go to infinity,
but in practice will be truncated at $n$.

We find that  the exact expression for the projected bispectrum $B_{\ell_1\ell_2\ell_3}$ is still given by equations
(\ref{eq:bisp1}) and  (\ref{eq.finalprojbisp}) where now the sum over $\ell$
goes up to $n$.

This, with equations (\ref{eq:bisp1}) and  (\ref{eq.finalprojbisp}), is the major new result of this paper. 

\subsection{Applications}
Equation (\ref{eq.finalprojbisp}) has therefore much wider applications
than the second-order gravitationally induced bispectrum considered so far.
The mathematics developed for this purpose can be straightforwardly applied to
CMB and gravitational lensing studies.  The
gravitational fluctuations and cosmological structures along the path of the
last-scattering surface photons. 
distort the CMB signal mainly through gravitational
lensing, the integrated Sachs-Wolfe effect (Sachs \& Wolfe 1967), the
Sunyaev-Zel'dovich effect (Sunyaev \& Zel'dovich 1980), and through the
Rees-Sciama effect (Rees \& Sciama 1968) and  other second order
effects (e.g. Mollerach \& Matarrese 1997 and references therein). In particular, if the primordial
fluctuations were Gaussian, many of these effects can introduce non gaussian features
in the CMB signal. The bispectrum is a powerful tool for detecting these
effects to probe the low-redshift Universe (Goldberg \& Spergel 1998, Spergel
\& Goldberg 1998).  Contributions to the CMB bispectrum induced
by secondary anisotropies during reionization (Cooray \& Hu 1999), non-linear
gravitational evolution (Luo \& Schramm 1994; Mollerach et al. 1995; Munshi,
Souradeep \& Starobinsky 1995) and  foregrounds
(e.g. Refregier, Spergel \& Herbig 1998) imprint specific signatures on the CMB
bispectrum which need to be subtracted from the signal in order to be able to test the gaussian nature of primordial
fluctuations.


On the other hand, primordial fluctuations can induce nonzero bispectrum in
the CMB, that encloses information about the physical mechanism that generated
them (e.g.  Falk, Rangarajan \& Srednicki 1993; Luo \& Schramm 1994, Gangui et
al. 1994, Mollerach et al. 1995, Gangui \& Mollerach 1996, Wang \&
Kamionkowski 1999, Gangui \& Martin 1999).  The evaluation of all these contributions to the observed
CMB bispectrum requires calculation of an integral as in our equation (5) and
(6), where $r^2 \psi(r)$ is replaced by an 
appropriate weight function.

In the local Universe, gravitational lensing provides a direct probe of the
mass fluctuations.
The study of Fourier space correlation functions of the gravitational weak
shear and convergence field is still in its infancy, but it is
potentially fruitful: it could give us detailed knowledge of the correlation
properties of the projected  mass distribution 
(e.g. Bernardeau et al. 1997 and references therein, Munshi 2000).

In the present paper we will use equation (\ref{eq.finalprojbisp}) together
with (\ref{eq:bisp1}) as an {\em exact} expression for the second-order perturbation theory bispectrum
of an angular catalogue with selection function $\psi(r)$ and galaxy power spectrum $P(k)$, assuming a 
local bias model with parameters
$b_1$ and $b_2$.  In principle, one can estimate the angular
bispectrum from a projected galaxy catalogue, and use likelihood
methods to constrain the bias parameters, which enter through the
$B_\ell$ terms.  In  Section 3, we compute the likely errors
from such a study, to determine if it is worthwhile to undertake such
an analysis with current catalogues.  Before we do so, it is worth
noting that, in the current form, it is very expensive to compute: in
the following subsection we rewrite it in a form more suitable for
practical evaluation.

\subsection{Practical evaluation of $B_{\ell_1\ell_2\ell_3}$}

From a computational point of view it is possible to speed up the
calculations considerably (and consequently make the problem
computationally manageable) by rewriting equation
(\ref{eq.finalprojbisp}) in terms of the function $\Theta^q_{\ell}$
defined as:
\be
\Theta_{\ell_i}^q(\ell_j,r)\equiv\int dk \Psi_{\ell_i}(k)k^2P(k)j_{\ell_j}(kr)k^q
\label{eq:theta}
\ee
where $q=-1,0,1$ and $\{i,j\}=\{1,6\}$ or $\{2,7\}$. This function can be
evaluated and tabulated in advance to speed up the analysis.
With this definition, we can write the components of the angular
bispectrum as
\ba
{\cal B}_{12}&=&\frac{1}{\overline{n}^3}\frac{16}{\pi}
\sqrt{\frac{(2\ell_1+1)(2\ell_2+1)(2\ell_3+1)}{(4\pi)^3}}i^{\ell_1+\ell_2}\times
\nn
& & \int dr_3 r_3^2\psi(r_3)\left(\frac{34}{21}c_1+c_2\right)\Theta^0_{\ell_1}(\ell_1,r_3)\Theta^0_{\ell_2}(\ell_2,r_3)
\left(^{\ell_1\;\;\;\ell_1\;\;\;0}_{0\;\;\;\;0\;\;\;\; 0} \right)
\left(^{\ell_2\;\;\;\ell_2\;\;\;0}_{0\;\;\;\; 0\;\;\;\; 0} \right)
\left(^{\ell_3\;\;\;\ell_1\;\;\;\ell_2}_{0\;\;\;\; 0\;\;\;\; 0} \right)
\left\{^{\ell_1\;\;\;\ell_2\;\;\;\ell_3}_{\ell_1\;\;\;\ell_2\;\;\;0}\right\}
+ \nn
& & c_1\sum_{^{\ell_1-1<\ell_6<\ell_1+1}_{\ell_2-1<\ell_7<\ell_2+1}}\left[\Theta_{\ell_1}^{-1}(\ell_6,r_3)\Theta_{\ell_2}^{+1}(\ell_7,r_3)+\Theta_{\ell_1}^{+1}(\ell_6,r_3)\Theta_{\ell_2}^{-1}(\ell_7,r_3)\right]
\left(^{\ell_1\;\;\;\ell_6\;\;\;\ell}_{0 \;\;\;\;0\;\;\;\; 0} \right)
\left(^{\ell_2\;\;\;\ell_7\;\;\;\ell}_{0\;\;\;\; 0\;\;\;\; 0} \right)
\left(^{\ell_3\;\;\;\ell_6\;\;\;\ell_7}_{0\;\;\;\; 0\;\;\;\; 0} \right)
\left\{^{\ell_1\;\;\;\ell_2\;\;\;\ell_3}_{\ell_7\;\;\;\ell_6\;\;\;\ell}\right\}
+\nn
& & \frac{8}{21}c_1\sum_{^{\ell_1-2<\ell_6<\ell_1+2}_{\ell_2-2<\ell_7<\ell_2+2}}
\Theta^0_{\ell_1}(\ell_6,r_3)\Theta^0_{\ell_2}(\ell_7,r_3)
\left(^{\ell_1\;\;\;\ell_6\;\;\;\ell}_{0 \;\;\;\;0\;\;\;\; 0} \right)
\left(^{\ell_2\;\;\;\ell_7\;\;\;\ell}_{0\;\;\;\; 0\;\;\;\; 0} \right)
\left(^{\ell_3\;\;\;\ell_6\;\;\;\ell_7}_{0\;\;\;\; 0\;\;\;\; 0} \right)
\left\{^{\ell_1\;\;\;\ell_2\;\;\;\ell_3}_{\ell_7\;\;\;\ell_6\;\;\;\ell}\right\}.
\label{eq.biscomputational}
\ea
Note that this analysis is appropriate for all-sky coverage, and
ignores shot noise.  This is a good approximation for the high
surface-density catalogues such as APM, in the range where
perturbation theory is valid. Estimators for noisy data and partial sky
coverage are presented in Heavens (1998) and Heavens (2000), see also \scite{ganguimartin00}.  Note also that numerical codes can run
into difficulties when computing the spherical harmonic expansion
and 3J symbols for high $\ell$. In Appendix C we give
asymptotic expressions at high $\ell$ for the 3J symbols that are
easily evaluated, and we present a way to calculate spherical
harmonics fast and accurately at high $\ell$.

\section{Error analysis for the bias parameter}

The spherical harmonic bispectrum in second-order perturbation theory
is a known function of the galaxy power spectrum, and depends on the
bias parameters $b_1$ and $b_2$ through $B_0, B_1, B_2$ (equation
\ref{eq.B0B1B2}).  Equation (\ref{eq.finalprojbisp}) relates therefore
two measurable quantities (the spherical harmonic bispectrum of
galaxies and the galaxy power spectrum) via the unknown bias
parameters $b_1$ and $b_2$.  The 3D power spectrum may be obtained
from the projected catalogue either by deconvolution of the angular
correlation function or the angular power spectrum
(e.g. \pcite{BE93,BE94}).  
In practice this is done in the
small-angle approximation.  In Appendix B we show that this is
perfectly adequate for the power spectrum.  We therefore have a full
prescription for the angular bispectrum in terms of observable
quantities and parameters which we wish to measure [equation
(\ref{eq.finalprojbisp}) or the computationally manageable equation
(\ref{eq.biscomputational})].  The problem is therefore suitable for a 
likelihood analysis to extract the bias parameter.   Such a programme
is a major undertaking, so it makes sense to compute the expected
error on the bias parameter first, to see whether the programme is likely to succeed.

\subsection{Likelihood analysis for $c_1$ and $c_2$}
We assume we have full sky coverage unless otherwise stated.
We define the quantity\footnote{{\bf Re}[x]
denotes the real part of the complex number x.} $d_{\alpha}$=${\bf
Re}[a_{\ell_1}^{m_1}a_{\ell_2}^{m_2}a_{\ell_3}^{m_3}]$
with $\ell_i$, $m_i$ such that $\left(^{\ell_1\;\;\;\ell_2\;\;\;\ell_3}_{m_1 m_2
m_3}\right)\neq 0$. For a given triplet
$\ell_1,\ell_2,\ell_3$ there are $(2\ell_1+1)(2\ell_2+1)$ distinct
$d_{\alpha}$.
From $d_{\alpha}$ we can build the unbiased estimator of
$B_{\ell_1\ell_2\ell_3}$,
$\hat{D}_{\alpha}=\frac{d_{\alpha}}{\left(^{\ell_1\;\;\;\ell_2\;\;\;\ell_3}_{m_1 m_2
m_3}\right)}$.

Since $\hat{D}_{\alpha}$ is unbiased, any combination
$D_{\alpha}=(\sum_{\alpha '}w_{\alpha '}\hat{D}_{\alpha '})/(\sum_{\alpha
'}w_{\alpha '})$,
where $w_{\alpha '}$ is a weight,
is
also an  unbiased estimator of $B_{\ell_1\ell_2\ell_3}$.

The optimum weight $w_{\alpha}$ that minimizes the variance
$\langle(D_{\alpha}-B_{\ell_1\,\ell_2\,\ell_3})^2 \rangle $ is
$w_{\alpha}=1/\sigma_{\hat{D}_{\alpha}}^2=\left(^{\ell_1\;\;\;\ell_2\;\;\;\ell_3}_{m_1\;
m_2\; m_3} \right)^2/\sigma^2_{d_{\alpha}}$ (cf Gangui \& Martin 2000).
The minimum variance estimator is
\be
D_{\alpha}=\frac{\sum_{m_1m_2m_3}\frac{
\left(^{\ell_1\;\;\;\ell_2\;\;\;\ell_3}_{m_1\;
m_2\; m_3}\right)d_{\alpha}}{\sigma^2_{d_{\alpha}}}}
{\sum_{m_1m_2 m_3}\frac{\left(^{\ell_1\;\;\;\ell_2\;\;\;\ell_3}_{m_1\;
m_2\; m_3} \right)^2}{\sigma^2_{d_{\alpha}}}}
\label{eq.estimatorB1}
\ee
The variance of  $d_{\alpha}$ does depend on $m$, but only weakly.
There is a leading term, independent of $m$, proportional to 3 angular
power spectra 
($C_{\ell_1}C_{\ell_2}C_{\ell_3}$), plus a sub-leading term proportional
to $B_{\ell_i\ell_j\ell_k}B_{\ell_{p}\ell_{q}\ell_{r}}\left(^{\ell_i\;\;\;\ell_j\;\;\;\ell_k}_{m_i\;
m_j\; m_k} \right)\left(^{\ell_{p}\;\;\;\ell_{q}\;\;\;\ell_{r}}_{m_{p}\;
m_{q}\; m_{r}} \right)$, where $\{i,j,k,p,q,r\}$ is a permutation of
$\{1,2,3,1,2,3\}$. If we ignore the $m$-dependence of this last 
term, then the estimator simplifies to
\be
D_{\alpha}=\sum_{m_i}d_{\alpha}\left(^{\ell_1\;\;\;\ell_2\;\;\;\ell_3}_{m_1\;
m_2\; m_3} \right), \mbox{ $i=1,2,3$}.
\label{eq.estimatorB2}
\ee
Strictly it is not the minimum variance estimator, but it is not far
from it, is much simpler, and is unbiased.

\subsection{A priori error for the bias parameter}

Since the quantity $\langle a_{\ell_1}^{m_1}a_{\ell_2}^{m_2}a_{\ell_3}^{m_3}
\rangle$ can be factorized as in equation (\ref{eq:2dbispluo}) it is 
possible to evaluate the expected error on $c_1$ estimation by
approximating the variance by its leading term, neglecting shot noise
and by considering uncorrelated data, obtaining:
\be
\sigma^{-2}_{c_1}=-\langle\frac{\partial^2{\cal L}}{\partial
c_1^2}\rangle
\simeq\sum_{\ell_i}\frac{B_{\ell_1\ell_2\ell_3}^2}{C_{\ell_1}C_{\ell_2}C_{\ell_3}}\sum_{m_i}\frac{\left(^{\ell_1\;\;\;\ell_2\;\;\;\ell_3}_{m_1\;m_2\;m_3}\right)^2}{N_{\ell_i}(m_i)}
\label{eq.sphharmapriorierror}
\ee
where ${\cal L}$ denotes the likelihood function.

The quantity $N_{\ell_i}(m_i)$ denotes the number of terms like
$C_{\ell_1}C_{\ell_2}C_{\ell_3}$ present in the covariance. It depends
on the configuration i.e. on the choice of the triplets of ${\ell}$'s.
It is useful to notice here that in the absence of $N_{\ell_i}(m_i)$ we have
\be
\sum_{m_1m_2m_3}\left(^{\ell_1\;\;\;\ell_2\;\;\;\ell_3}_{m_1\;m_2\;m_3}
\right)^2=1.
\ee
Equation (\ref{eq.sphharmapriorierror}) assumes that the covariance
matrix is diagonal, this means that different bispectrum estimators
$\langle a_{\ell_1}^{m_1}a_{\ell_2}^{m_2} a_{\ell_3}^{m_3}
\rangle$ are uncorrelated. In the case of a survey with full sky coverage,
similarly to the three-dimensional case treated in MVH97, the
covariance matrix is well approximated by a diagonal matrix if each
$a_{\ell}^m$ appears in one estimator only.  However,
in the presence of a mask, different $a_{\ell}^m$ are
correlated, therefore (\ref{eq.sphharmapriorierror}) might no
longer be valid.

For an order-of-magnitude estimation of the expected error on the bias
parameter, let us consider only equilateral configurations
(i.e. configurations where $\ell_1=\ell_2=\ell_3$), and assume full
sky coverage for a survey with the APM selection function. It is easy to estimate the error achievable on $c_1$
using equation (\ref{eq.sphharmapriorierror}) and considering that
second order perturbation theory should hold up to $\ell=35$. This
choice is justified by the following argument: in the
three-dimensional galaxy distribution, second-order perturbation
theory breaks down at $k\sim 0.6$ (Mpc $h^{-1}$)$^{-1}$ (cf
 MVH97, although it depends on the power spectrum slope and can be
smaller, see e.g. \scite{SCFFHM98}) that
corresponds to a scale of the order of 10 Mpc $h^{-1}$. At the medium
depth of the APM survey (335 Mpc $h^{-1}$), this subtends an angle of
about 0.03 radians (in agreement with the findings of \pcite{GB98}),
corresponding with $\ell \sim 33$.  This order of magnitude
calculation yields an estimate for the error on $c_1$ of about $\pm
3.5$, which is not really encouraging.  However, this is only an
order-of magnitude calculation: a more rigorous treatment
is implemented in the next section.

\subsection{The choice of the triplets}

As already discussed in MVH97, the choice of the triplets to evaluate
the bispectrum is very wide, but, speed and memory considerations
force one to simplify the analysis by ensuring that the covariance
matrix is diagonal, for a full sky survey, this can be achieved by
ensuring that each $\ell$ appears only in one triplet. 
The choice of
the ratio between the $\ell$'s (the shape) of a triplet, is influenced
by the behavior of the bispectrum: triplets with the same shape give
an almost degenerate information on $c_1$ and $c_2$, in practice each
shape can constrain a linear combination of $c_1$ and $c_2$: the
likelihood will be aligned along a straight line in the $c_1$,$c_2$
plane.  The best choice to try to lift this additional degeneracy is
to combine the likelihood for equilateral triplets
($\ell_1=\ell_2=\ell_3$) with the likelihood for degenerate triplets
($\ell_1=\ell_2$ and $\ell_3=2\ell_1$).

\subsection{Covariance}
To perform a likelihood analysis we need an expression for the covariance
matrix for our estimator $\widehat{B}_{\ell_1\ell_2\ell_3}$.
It is easy to verify that, if each $\ell$ appears only in one
$\widehat{B}_{\ell_1\ell_2\ell_3}$ then the
$\widehat{B}_{\ell_1\ell_2\ell_3}$ are uncorrelated, that is:
\be
\langle
\widehat{B}_{\ell_1\ell_2\ell_3}\widehat{B}_{\ell_4\ell_5\ell_6}\rangle=\langle\widehat{B}_{\ell_1\ell_2\ell_3}\rangle
\langle\widehat{B}_{\ell_4\ell_5\ell_6} \rangle
\ee
if $\ell_i\neq \ell_j$, where $i=1,2,3$ and $j=4,5,6$. This means that
the off-diagonal terms of the covariance matrix are zero.  In fact:
\be
\langle
\widehat{B}_{\ell_1\ell_2\ell_3}\widehat{B}_{\ell_4\ell_5\ell_6}\rangle=
\sum_{m_1m_2m_3}\sum_{m_4m_5m_6}\langle a_{\ell_1}^{m_1}a_{\ell_2}^{m_2}a_{\ell_3}^{m_3}a_{\ell_4}^{m_4}a_{\ell_5}^{m_5}a_{\ell_6}^{m_6}
\rangle\ \left(^{\ell_1\;\;\;\ell_2\;\;\;\ell_3}_{m_1\;m_2\;m_3} \right) \left(^{\ell_4\;\;\;\ell_5\;\;\;\ell_6}_{m_4\;m_5\;m_6} \right)
\label{eq.cov}
\ee
where we used the fact that $a_{\ell}^{*m}=(-1)^m a_{\ell}^{-m}$.
Analogously to MVH97, the quantity $\langle
a_{\ell_1}^{m_1}a_{\ell_2}^{m_2}a_{\ell_3}^{m_3}a_{\ell_4}^{m_4}a_{\ell_5}^{m_5}a_{\ell_6}^{m_6}\rangle$
can be split into:
\begin{itemize}
\item[a)] 15 cyclical permutations of the kind $\langle a_{\ell_1}^{m_1}a_{\ell_2}^{m_2}\rangle
\langle a_{\ell_3}^{m_3}a_{\ell_4}^{m_4}\rangle
\langle a_{\ell_5}^{m_5}a_{\ell_6}^{m_6}\rangle$  that are all zero if
$\ell_i\neq \ell_j$, where  $i=1,2,3$ and $j=4,5,6$,

\item[b)] 1 term
$$B_{\ell_1\ell_2\ell_3}B_{\ell_4\ell_5\ell_6}\sum_{m_1m_2m_3}\sum_{m_4m_5m_6}\left(
^{\ell_1\;\;\;\ell_2\;\;\;\ell_3}_{m_1\;m_2\;m_3}\right)^2
\left(^{\ell_4\;\;\;\ell_5\;\;\;\ell_6}_{m_4\;m_5\;m_6} \right)^2 =
\langle\widehat{B}_{\ell_1\ell_2\ell_3} \rangle
\langle\widehat{B}_{\ell_4\ell_5\ell_6} \rangle$$

\item[c)] 9 cyclical permutations of the kind:
$$B_{\ell_i\ell_j^{\prime}\ell_k}B_{\ell_i^{\prime}\ell_j\ell_k^{\prime}}\sum_{m_im_jm_k}\sum_{m_i^{\prime}m_j^{\prime}m_k^{\prime}}
\left(_{m_i\; m_j^{\prime}\;m_k}^{\ell_i\;\;\;\ell_j^{\prime}\;\;\;\ell_k}\right)
\left(^{\ell_i^{\prime}\;\;\;\ell_j\;\;\;\ell_k^{\prime}}_{m_i^{\prime}\;m_j\; m_k^{\prime}}\right)
\left(^{\ell_1\;\;\;\ell_2\;\;\;\ell_3}_{m_1\;m_2\;m_3}\right)
\left(_{m_4\;m_5\;m_6}^{\ell_4\;\;\;\ell_5\;\;\;\ell_6}\right)
$$
where $i,j,k$ is any permutation of 1,2,3 and $i^{\prime}j^{\prime}k^{\prime}$
denotes any permutation of 4,5,6.
These terms in c) are all zero unless there are repeated $\ell$ in two different $D_{\alpha}$.
\end{itemize}
The term in b) cancels when subtracting the mean to obtain the covariance.
Let us now consider the diagonal terms of the covariance matrix: these are
given by equation  (\ref{eq.cov}) with the following identities for the indices:
$1=4$, $2=5$, $3=6$.
For symmetry considerations we can restrict ourselves to consider
$\ell_1\le\ell_2\le \ell_3$.
In the case where $\ell_1<\ell_2<\ell_3$ we have:
\begin{itemize}
\item[] in a) only one term surviving, giving $C_{\ell_1}C_{\ell_2}C_{\ell_3}$.
\item[] in c) using (\ref{eq.orth2}), $3B_{\ell_1\ell_2\ell_3}^2$
\end{itemize}

In the case where $ \ell_1=\ell_2<\ell_3$ we have:
\begin{itemize}
\item[ ]in a) $C_{\ell_1}^2C_{\ell_3}\left[2+\sum_{m
m^{\prime}}\left(^{\ell_1\;\;\;\ell_1\;\;\;\ell_3}_{m\;-m
\;0}\right)\left(^{\ell_1\;\;\;\ell_1\;\;\;\ell_3}_{m^{\prime}\;-m^{\prime}\;0}\right) \right]$
in the particular case where $\ell_3=2\ell_1$ (degenerate configurations) as
shown in Appendix C (eq. \ref{eq:3j3jll2l}) we obtain to very good
approximation: $C_{\ell_1}^2C_{\ell_3}\left[2+\sqrt{(2\pi
\ell_1)}/(1+4\ell_1)\right]$
\item[ ] in c) $5 B_{\ell_1\ell_1\ell_3}^2$
\end{itemize}

For equilateral configurations where $\ell_1=\ell_2=\ell_3=\ell$ we have:
\begin{itemize}
\item[]in a) $C_{\ell}^3\left[6+9\sum_{m m^{\prime}}\left(^{\ell\;\;\;\; \ell\;\;\;\; \ell}_{m
-m\; 0}\right)\left(^{\ell\;\;\;\; \ell \;\;\;\;\ell}_{m^{\prime} -m^{\prime}\; 0 }\right) \right]$
which, to a very good approximation, is $\simeq C_{\ell}^3\left[6+9\times 1.15/(2\ell+1)\right]$
\item[ ]in c) $9 B_{\ell\ell\ell}^2$
\end{itemize}

\subsection{Likelihood analysis of a simulated catalogue}
We created an all-sky catalogue with the APM selection function
$\phi(r)\propto r^{-0.1}\exp[(-r/335)^2]$ (e.g. \scite{apmselfn95},
\scite{GB98}), by replicating an N-body simulation of 500
Mpc $h^{-1}$ side and then sparsely sampling and projecting to the
plane of the sky 2303636 particles (galaxies for our purposes)
accordingly to the APM selection function.  The simulation was
supplied by J. Peacock using the AP$^3$M code (\pcite{C91}). The
characteristics are: $128^3$ particles, CDM transfer function
(\pcite{BBKS}), $\Gamma=0.25$, $\Omega=0.3$, $\Lambda=0.7$, evolved to
$\sigma_8$=1 to best fit the observed cluster abundance; this 
choice gives also a good fit to the COBE 4-year data (e.g. \pcite{Teg96}).  The rate of
sampling used ensures that the probability of selecting the same
particle in the simulation more than once in the replicated box is
negligible.

\begin{figure}
\begin{center}
\setlength{\unitlength}{1mm}
\begin{picture}(90,70)
\includegraphics{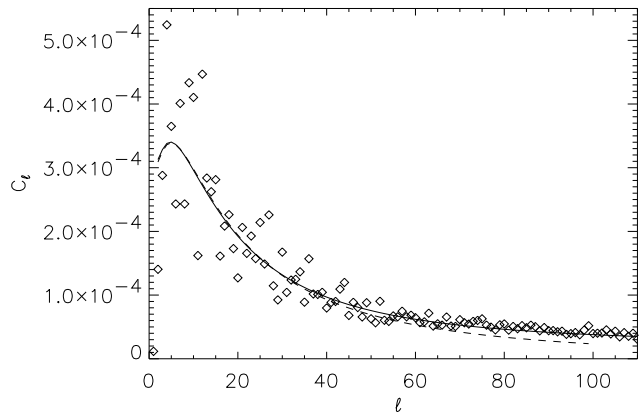}
\end{picture}
\end{center}
\caption{The power spectrum in spherical harmonics for the simulated projected
catalogue. The points are the $C_{\ell}$ measured for the catalogue, the solid
line is the underlying power spectrum, and the dashed line is the underlying
linear power spectrum. The underlying power spectrum is obtained from the 3D
one by applying the selection function and the full-treatment projection. Deviations from linear theory are already evident at
$\ell \sim 40$.}
\label{fig.clsim}
\end{figure}

\begin{figure}
\begin{center}
\setlength{\unitlength}{1mm}
\begin{picture}(90,70)
\includegraphics{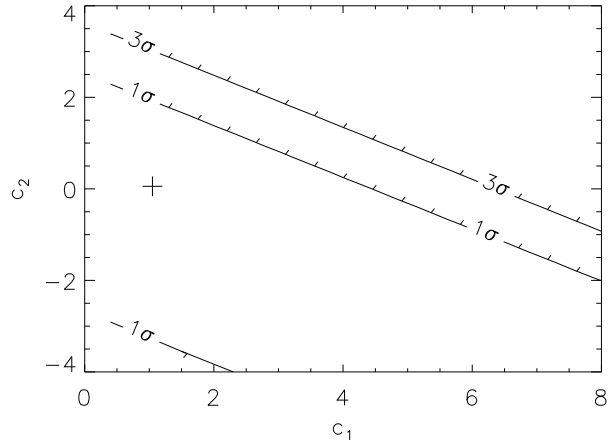}
\end{picture}
\end{center}
\caption{Likelihood contours for equilateral triplets
configurations. The two levels (where the downhill direction is indicated for
clarity) are the 1$-\sigma$ and 3-$\sigma$ confidence
levels and the $+$ indicates where the true value for the parameters
lies. Perturbation theory breaks down at $\ell \sim 40$ here $\ell$ up to 42
are considered even though the likelihood only for the triplets between
$\ell=40$ and $\ell=42$ includes the true value just within the boundary of
the $3-\sigma$ level. This configuration does not place strong constraints on
the bias parameter.}
\label{fig.likeq}
\end{figure}

\begin{figure}
\begin{center}
\setlength{\unitlength}{1mm}
\begin{picture}(90,70)
\includegraphics{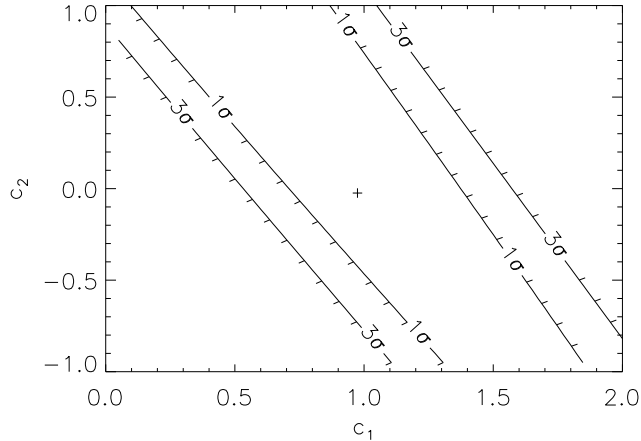}
\end{picture}
\end{center}
\caption{Likelihood contours for degenerate triplets
configurations. The two levels are the 1$-\sigma$ and 3-$\sigma$ confidence
levels and the $+$ indicates where the true value for the parameters
lies. Perturbation theory breaks down at $\ell_{\rm short}=40$.}
\label{fig.likdeg}
\end{figure}

Figure \ref{fig.clsim} shows the angular power spectrum for the
simulated projected catalogue. The solid line is the underlying 
power spectrum obtained from the 3D one by applying  convolution with the
selection function and full-treatment projection; the dashed line is the
underlying linear power spectrum. Deviations for
linear theory are already evident at $\ell \sim 40$.
In order to lift the degeneracy between $c_1$ and $c_2$ we consider
equilateral and degenerate configurations.
The likelihood for equilateral configurations is shown in Figure
\ref{fig.likeq}; it does not give a strong constraint
on the bias: perturbation theory breaks down at $\ell \sim 40$: up to $\ell
< 40$ there are only 18 independent equilateral triplets\footnote{18 is the
number of independent $D_{\alpha}$ as defined in eq. 35 with $\ell_i=\ell<40$, $i=1,2,3$. The
presence of the 3-J symbol for equilateral configurations requires $\ell$ to
be even, $\ell=2$ is discarded because it would be  contaminated by the galaxy
quadrupole.}.
The likelihood for degenerate configurations gives a better constraint on the
bias. Perturbation theory for this configuration breaks down where the short
$\ell$ is
$\ell=40$. Likelihood contours for degenerate triplets configurations where
the short $\ell$ is $20< \ell\leq 40$, are shown
in Figure \ref{fig.likdeg}.
Since the likelihood for equilateral configurations is quite broad, even
combining it to the likelihood for degenerate configurations does not modify Figure \ref{fig.likdeg} sensibly.

From this analysis we can conclude that from a two-dimensional galaxy  survey
with the APM selection function, even if it is an all-sky survey, it is only
possible to constrain a combination of the linear and quadratic bias
parameter, or, alternatively, if we {\em assume} (without
justification) that the bias is linear (i.e. $b_2=0$), 
$0.7<c_1<1.4$ or  $0.7<b_1<1.4$, at  68\% confidence. 

\section{Conclusions}

In this paper, we have presented the formalism for translating the 3D
bispectrum of a sample population into the angular bispectrum in spherical
harmonics. As discussed in section 2.1, this method has applications in a
variety of areas, such as microwave background studies, gravitational lensing
and analysis of angular galaxy catalogues. We have investigated the last of these in
detail in this paper: since the bispectrum is a measurable quantity, and
its theoretical expression depends on measurable quantities via the
unknown bias parameter, it is possible to extract the bias parameter
via a likelihood analysis.  We have therefore investigated its use as
a tool for measuring the bias parameter for projected galaxy surveys.  In
principle, it is an alternative method to using 3D galaxy redshift
surveys, without the complicating effects of redshift distortions and
higher shot noise. 
Recently, other methods based on second-order perturbation
theory have been proposed to measure the bias parameter from 2D galaxy
catalogues (see e.g. \pcite{FG99,FryThomas99}).
Frieman \& Gaztanaga (1999) studied the reduced 3-point correlation
function on the sky.  The error analysis in real space is more
complicated because of strong correlations between the estimates.
Frieman \& Gaztanaga conclude that $b_1 \ll 1.5$ or so, giving a
comparable error to our analysis (section 3.5) if $b_2$ is assumed to
be zero.  We emphasize that allowing a non-zero quadratic bias term
opens up a wide range of acceptable linear bias parameters.

The analysis of Fry \& Thomas (1999) is closest to ours.  They consider
the bispectrum, but present results in the small-angle approximation
only.  They do go some way in writing down the general expression for
the angular bispectrum in spherical harmonics in terms of the 3D
bispectrum.  In this paper, by expanding the (general) dependence of
the 3D bispectrum on angle between wavevectors in Legendre
polynomials, we were able to derive a practical general relationship
which is computable with few numerical integrations.

We have 
calculated the expected
error on the linear bias parameter from an all-sky catalogue with a
selection function similar to the APM survey.  We find that the
results are not encouraging for projected catalogues, and that it is
preferable to undertake a bispectrum study of 3D galaxy redshift
surveys such as the AAT 2dF or the Sloan Digital Sky Survey, using the
methods discussed in MVH97 and \scite{VHMM98}.  Tests on simulated
projected catalogues confirm our analytic findings. In a similar way as for 3D
surveys (see discussion in MVH97), one can reduce the
errors by subdividing the sky \footnote{The procedure of subdivision to increase the S/N appears
counter-intuitive.  In fact, nothing more is gained by this than by
relaxing the precise shape of the triangles.  It is easiest to
demonstrate this in Fourier space in 3D.  MVH97 considered the
bispectrum $B({\bf k}_1, {\bf k}_2, {\bf k}_3)$ and demonstrated that
the S/N increases $\propto N^{1/2}$, where $N$ is the number of
subvolumes.  Alternatively, in increasing the volume by a factor $N$,
one has more triangles to analyse.  MVH97's analysis includes the
density of ${\bf k}_1$ states;  relaxing the triangle shape
configuration increases the number of ${\bf k}_2$ states by the ratio
of the density of states (i.e. $N$).  ${\bf k}_3$ is fixed at $-{\bf
k}_1-{\bf k}_2$, so the number of triangles for given ${\bf k_1}$ is
$\propto N$, giving the same increase of S/N (see Verde 2000).  In practice
correlations may modify the details.  This argument is a variant of
that in Press et al. (1992).}, but by modest factors which do not
change this basic conclusion, and at the expense of considerable
complication in the analysis through window convolutions.

\section*{Acknowledgments}
LV acknowledges the support of a TMR grant.
We thank John Peacock for providing the N-body simulation, Marc
Kamionkowski, Ari Buchalter, Silvia Mollerach and Andy Taylor for useful
discussions and the anonymous referee for helpful comments.

\section*{Appendix A: Spherical harmonics}

Spherical harmonics are the natural basis for describing a
two-dimensional random field on the sky.  The definition we use here
is expressed in terms of the associate Legendre functions
$P_{\ell}^{m}(\cos\theta)$:
\be
Y_{\ell}^m(\Omega)\equiv
Y_{\ell}^m(\theta,\phi)=\sqrt{\frac{(2\ell+1)(\ell-m)!}{4\pi(\ell+m)!}}P^m_{\ell}(\cos\theta)e^{\imath
m\phi}\times\left\{^{(-1)^m\mbox{  for $m\geq 0$}}_{1 \mbox{   for $m < 0$}}\right.
\ee
where $\ell$ and $m$ are integers and $\ell\geq0$, $-\ell<m<\ell$.
Their orthogonality relation is:
\be
\int_{4 \pi}d\Omega Y_{\ell_1}^{m_1}(\Omega)
Y_{\ell_2}^{*m_2}(\Omega)=\delta^K_{\ell_1\ell_2}\delta^K_{m_1m_2}.
\label{eq.spharmorth}
\ee
Any two dimensional pattern  $f(\Omega)$ on the surface of a sphere can be
expanded as:
\be
f(\Omega)=\sum_{\ell m}a_{\ell}^{m}Y_{\ell}^{m*}(\Omega)
\ee
where
\be
a_{\ell}^{m}=\int d\Omega Y^{m}_{\ell}(\Omega)f(\Omega).
\ee

Useful  relations involving the spherical harmonics are:
\be
Y_{\ell}^{m*}(\Omega)=(-1)^mY^{-m}_\ell(\Omega)\mbox{  ,  }
Y_{\ell}^{m}(-\Omega)=(-1)^{m+\ell}Y^{m}_\ell(\Omega).
\label{eq.spharmconj}
\ee
and the identities:
\be
\sum_{\ell m}Y_{\ell}^{*m}(\Omega)Y_{\ell}^{m}(\Omega^{\prime})=\delta(\Omega-\Omega^{\prime})
\ee

\be
\exp(i \vk\cdot\vr)=4\pi\sum_{\ell
m}i^{\ell}j_{\ell}(kr)Y_{\ell}^{*m}(\Omega_\vk)Y_{\ell}^{m}(\Omega_\vr)
\label{eq.expspharm}
\ee
\be
P_{\ell}(\cos\theta)=\frac{4\pi}{2\ell+1}\sum_{m=-\ell}^{\ell}Y_{\ell}^{m}(\Omega_{\vr})Y^{*m}_{\ell}(\Omega_{\vk})
\label{eq.legendrep}
\ee
where $\theta$ denotes the angle between the vectors $\vr$ and $\vk$.
The latter is the addition theorem for spherical harmonics.

\section*{Appendix B: the power spectrum and the small-angle approximation}

The power spectrum of  a 2-D distribution on the plane on the sky is
given by the set of $C_{\ell}$
defined as:
\be
\langle a_{\ell}^m a_{\ell^{\prime}}^{m^{\prime}*} \rangle=C_{\ell} 
\delta^{K}_{\ell\ell^{\prime}}\delta^{K}_{m m^{\prime}}.
\ee
The corresponding angular two-point correlation function
can be expanded as
\be
C(\theta)=\frac{1}{4\pi}\sum_{\ell}(2\ell+1)C_{\ell}P_{\ell}(\cos\theta)
\label{eq.ctheta}
\ee
with inverse relation is:
\be
C_{\ell}=2\pi \int_{-1}^{1}C(\theta)P_{\ell}(\cos\theta)d\cos(\theta)
\ee
On the other hand, for a plane two-dimensional distribution with 
power spectrum $P_{2D}(\kappa)$
the two-point correlation function is:
\be
 w(\theta)=\frac{1}{(2\pi)^2}\int P_{2D}(\kappa)\exp(i {\vkappa} \cdot {\vtheta})d^2{\bf
\kappa}=
\frac{1}{(2\pi)^2}\int_{0}^{\infty}\int_{0}^{2\pi}P_{2D}(\kappa)\cos(\kappa\theta\cos\phi)d\phi
\kappa d\kappa=\frac{1}{(2\pi)}\int_0^{\infty}P_{2D}(\kappa) J_0(\kappa\theta) \kappa
d\kappa 
\ee
In the small angle approximation, i.e. in equation (\ref{eq.ctheta})
for small $\theta$, we have that $P_{\ell}(\cos\theta)\sim
J_0[(\ell+1/2)\theta]$, but, since small angular patches restrict us
to high $\ell$, $P_{\ell}(\cos\theta)\sim J_0(\ell \theta)$.  We can
therefore conclude that in the small-angle approximation
\be
\ell\longrightarrow \kappa\mbox{ ;        }
\;\;\;\;C_{\ell}\longrightarrow P_{2D}({\ell})
\label{eq.smallanglemap}
\ee
For a real angular catalogue the two-dimensional galaxy density in the
sky is obtained as follow.  Let the true three-dimensional galaxy
density field be $\rho(\vr)$ and the selection function be $\psi(r)$,
normalized here to $\int dr r^2 \psi(r)=1$. It is straightforward to
obtain an expression for the angular power spectrum given the
three-dimensional one:
\be
\langle a_{\ell_1}^{m_1}a_{\ell_2}^{m_2*}\rangle =\left\{\begin{array}{ll}
\frac{1}{\overline{n}^2}\frac{2}{\pi}\int dk k^2 P(k)\left[\int dr
r^2 \psi(r)j_{\ell_1}(k_1 r)\right]^2 & \mbox{if $m_1=-m_2$ and $\ell_1=\ell_2$}\\
0 &\mbox{otherwise}
\end{array}
\right.
\label{eq.projpsexact}
\ee
In the small-angle approximation this is (\pcite{Kaiser92}, \pcite{BKJ99}):
\be
P_{2D}(\kappa)=\int_{0}^{\infty}dr P_{3D}(\kappa/r)\psi^2(r)r^2
\label{eq.projpssmallangle}
\ee
Also in the presence of the selection function we can check that mapping
(\ref{eq.smallanglemap}) is valid and we can asses the limit of validity for
the small angle approximation for the power spectrum.

\begin{figure}
\begin{center}
\setlength{\unitlength}{1mm}
\begin{picture}(90,70)
\includegraphics{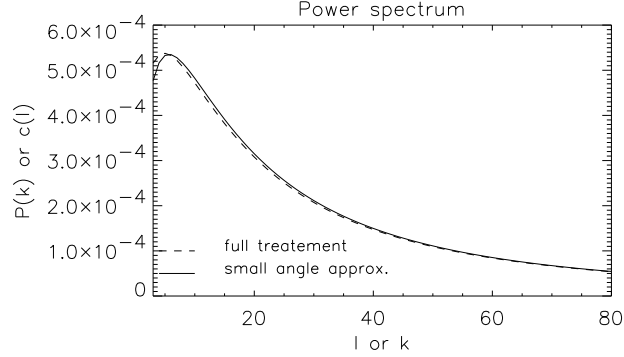}
\end{picture}
\end{center}
\label{fig.smallangleps}
\caption{The small-angle approximation for the angular power spectrum works
very well for $\ell > 20$ introducing an error smaller that 3\%.}
\end{figure}

Assuming the APM selection function $\phi(r)\propto
r^{-0.1}\exp[(-r/335)^2]$, and a CDM power spectrum (\pcite{EBW92})
with $\Gamma= 0.25$, we compared the angular power spectrum obtained
with the exact projection as in equation (\ref{eq.projpsexact}) and
with the small angle approximation [equation
\ref{eq.projpssmallangle}]. The result is shown in Figure
\ref{fig.smallangleps}: the small angle approximation introduces an
error smaller than 3\% for $\ell > 20$.


\section*{Appendix C: Useful formulae for the high $\ell$ regime.}

Library routines dealing with spherical harmonics at high $\ell$ can
sometimes fail.  In this appendix, we present some asymptotic results
which can avoid problems.

The 3J symbol $\left(^{\ell_1\ell_2\ell_3}_{0\;\;0\;\;0}\right)$ may be written
as:
\be
\left(^{\ell_1\ell_2\ell_3}_{0\;\;0\;\;0}\right)=(-1)^L
\sqrt{\frac{(L-2\ell_1)!(L-2\ell_2)!(L-2\ell_3)!}{(L+1)!}}
\frac{(L/2)!}{(L/2-\ell_1)!(L/2-\ell_2)!(L/2-\ell_3)!}
\label{eq:clebshzero}
\ee
where $L=\ell_1+\ell_2+\ell_3$.

In the special case where $\ell_1=\ell_2=\ell$ and $\ell_3=2\ell$ this becomes:
\be
\frac{2[\Gamma(2\ell)]^2}{\sqrt{\ell} \sqrt{1+4\ell}[\Gamma(\ell)]^2\sqrt{\Gamma(4\ell)}}
\label{eq:clebshzerodeg}
\ee

Numerical routines to calculate the 3J symbols usually encounter
problems for large ${\ell}$. We therefore evaluated an approximation
based on the Stirling approximation: i.e.  $n!=\Gamma(n+1)$ and
(\pcite{GR65})
\be
\Gamma(z)\sim e^{-z}z^{z-1/2}(2\pi)^{1/2} \;\;\;\mbox{  for large }\;\;\; z.
\label{eq:gammastirling}
\ee
This approximation for the $\Gamma$ function is quite good, in fact it
introduces an error of only 4\% at $z=2$. Using this approximation we
obtain for eq.  (\ref{eq:clebshzero}):
\be
\left(^{\ell_1\ell_2
\ell_3}_{0\;\;0\;\;0}\right)\longrightarrow \frac{(-1)^{L/2}}{\sqrt{2\pi}}
\frac{2^{1/4}\sqrt{e}}{(L+1)^{L/2+1/2+1/4}}\frac{L^{L/2+1/2}}{[(L/2-\ell_1)(L/2-\ell_2)(L/2-\ell_3)]^{1/4}}.
\ee
This approximation introduces a small error of a few percent at
${\ell}\sim 20$.  For equation (\ref{eq:clebshzerodeg}) we obtain:
\be
\left(^{\ell\;\;\ell\;\;
2\ell}_{0\;\;0\;\;0}\right)\longrightarrow\frac{1}{\sqrt{1+4\ell}}\left(\frac{4}{\ell 2 \pi}\right)^{0.25}
\ee
When the $m$'s are non-zero it is still possible to find a simple expression
for the 3J symbol in special cases. For example if
$\ell_3=\ell_1+\ell_2$ we have:
\be
\left(^{\ell_1 \;\;\ell_2\;\;\;\;\; \ell_1+\ell_2}_{m_1 m_2 -m_1-m_2}\right)=(-1)^{\ell_1-\ell_2+m_1+m_2}\sqrt{\frac{(2\ell_1)!(2\ell_2)!(\ell_1+\ell_2+m_1+m_2)!(\ell_1+\ell_2-m_1-m_2)!}{(2\ell_1+2\ell_2+1)!(\ell_1+m_1)!(\ell_2+m_2)!(\ell_1-m_1)!(\ell_2-m_2)!}}
\ee
In the special case where $\ell_1=\ell_2=\ell$ the previous expression can be further
simplified and approximated --using (\ref{eq:gammastirling})-- by:
\be
\left(^{\ell \;\;\;\;\ell\;\;\;\;\;\;\;\; 2\ell}_{m_1 m_2
-m_1-m_2}\right)\longrightarrow(-1)^{m_1+m_2}\frac{(\ell 2 \pi)^{1/4}}{2^{2\ell}\sqrt{4\ell+1}}\sqrt{\frac{(2\ell+m_1+m_2)!(2\ell-m_1-m_2)!}{(l+m_1)!(\ell+m_2)!(\ell-m_1)!(\ell-m_2)!}}
\ee
In the calculation of the covariance matrix for ``degenerate''configurations
for the quantities $B_{\ell\ell2\ell}$ we came across with  the following  sum
over $m$ of  a
product of two 3J-symbols. An useful expression for it is
the following:
\be
\sum_{m_1,m_2}\left(^{\ell\;\;\;\;\;\;\ell\;\;\;\;\; 2\ell}_{m_1\;-m_1\;0}\right)\left(^{\ell\;\;\;\;\;\;\ell\;\;\;\;\;
2\ell}_{m_2\;-m_2\;0}\right)=\frac{2^{(2+4\ell)}\ell^3\Gamma(2\ell)^4}{(1+4\ell)\Gamma(4\ell)\Gamma(2\ell+1)^2}
\equiv\frac{2^{4\ell}\ell
[(2\ell-1)!]^2}{(1+4\ell)(4\ell-1)!}.
\label{eq:3j3jll2l}
\ee

For large $\ell$ using the above approximation for the Gamma function we
obtain:
\be
\frac{2^{4\ell}\ell[\Gamma(2\ell)]^2}{(1+4\ell)\Gamma(4\ell)}\longrightarrow\frac{\sqrt{2\pi}\sqrt{\ell}}{(1+4\ell)}
\mbox{  (for large $\ell$)}
\ee
this approximation works very well also at low $\ell$, in fact
introduces an error below 1\% at $\ell=6$.

The orthogonality relations for 3J-symbols are also widely used:
\be
\sum_{m_1\;m_2\;m_3}\left(^{\ell_1\;\;\;\ell_2\;\;\;\ell_3}_{m1\;\;m2\;\;m3}\right)^2=1
\label{eq.orth1}
\ee
and
\be
\sum_{m_1\;m_2}\left(^{\ell_1\;\;\;\ell_2\;\;\;\ell_3}_{m1\;\;m2\;\;m3}\right)\left(^{\ell_1\;\;\;\ell_2\;\;\;\ell_4}_{m1\;\;m2\;\;m4}\right)=\frac{1}{(2
\ell_3+1)}\delta^K_{\ell_3\;\ell_4}\delta^K_{m_3\;m_4}
\label{eq.orth2}
\ee
When calculating the spherical harmonic coefficients for a galaxy distribution
on the celestial sphere, numerical problems arise with the
associate Legendre polynomials at high $\ell$. Routines based on the  
recurrence relations used for
example by the numerical recipes routines (\pcite{NUMREC})  
fails at $\ell \sim 35$ if $m \sim
\ell$, at slightly higher $\ell$ for $m \ll l$.
The asymptotic expansion for the associate Legendre polynomial
(e.g.\pcite{GR65}):
\be
P_{\ell}^{m}(\cos\theta)\simeq
\frac{2}{\sqrt{\pi}}\frac{\Gamma(\ell+m+1)}{\Gamma(\ell+3/2)}\frac{\cos[(\ell+1/2)\theta-\pi/4+m\pi/2]}{\sqrt{2\sin\theta}}+{\cal O}(1/\ell)
\ee
 is valid for $\mid \ell\mid \gg\mid m\mid$, $\mid \ell \mid \gg 1$ and
 $\epsilon<\theta<\pi-\epsilon$.

The use of the  recursive relations  involve the calculation of
$(2\ell-1)!!$ that at high $\ell$  can create numerical problems.
Using the following expression for the $\Gamma$ function:
\be
\Gamma(n+1/2)=\frac{\sqrt{\pi}}{2^n}(2n-1)!!
\ee
and (\ref{eq:gammastirling}) for the $\Gamma$ function for big argument we
obtain:

\be
(2n-1)!!=\frac{2^{n+1/2}(n+1/2)^n}{e^{n+1/2}}
\ee

This approximation introduces an error of only a few percent for $n \sim
10$.

When calculating the spherical harmonics at higher $\ell$, problems arise not
only with the associated Legendre polynomials, but also with the part that
involves the ratio of two factorials.
A better way to calculate the spherical harmonics, fast and accurate to high
$\ell$ is based on the algorithm proposed by \cite{Muciacciaetal97}. In
essence the numerical problems can be avoided by defining the normalized
associated Legendre polynomials $\lambda_{\ell}^m$:
\begin{equation}
\lambda_{\ell}^{m}(\cos\theta)\equiv\sqrt{\frac{2\ell+1}{4\pi}\frac{(\ell-m)!}{(\ell+m)!}}P_{\ell}^m(\cos\theta)
\end{equation}
The recurrence relation for $\lambda_{\ell}^m$ is:
\be
\lambda_{\ell}^m(x)=\left[x\lambda_{\ell-1}^m(x)-\sqrt{\frac{(\ell+m-1)(\ell-m-1)}{(2\ell-3)(2\ell-1)}}\lambda_{\ell-2}^m(x)\right]\sqrt{\frac{4\ell^2-1}{\ell^2-m^2}}
\ee
with expressions for the  starting values:
\be
\lambda_m^m(x)=(-1)^m\sqrt{\frac{2m+1}{4\pi}}\frac{(2m-1)!!}{\sqrt{(2m)!}}(1-x)^{m/2}
\label{eq:lmm}
\ee
\be
\lambda_{m+1}^m(x)=x\sqrt{2m+3}\lambda_m^m(x)
\ee
numerical evaluation can be greatly speeded up by noticing that the factor in
(\ref{eq:lmm}) depends only on $m$ and can therefore be calculated and/or
tabulated for $m\leq \ell_{max}$ only once.

\end{document}